\documentclass[%
 reprint,
superscriptaddress,
showpacs,
 amsmath,amssymb,
 aps,prl
floatfix,
]{revtex4-1}
\usepackage{mwe}
\usepackage{graphicx}
\usepackage{dcolumn}
\usepackage{bm}
\usepackage{hyperref}
\usepackage{xcolor}
\usepackage{color}
\usepackage[normalem]{ulem}

\usepackage{ulem}

\begin{document}



\title{Interdependent scaling exponents in the human brain}

\author{Daniel M. Castro}
\affiliation{Departamento de F{\'\i}sica, Centro de Ci\^encias Exatas e da Natureza, Universidade Federal de Pernambuco,
Recife, PE, 50670-901, Brazil}

\author{Ernesto P. Raposo}
\affiliation{Departamento de F{\'\i}sica, Centro de Ci\^encias Exatas e da Natureza, Universidade Federal de Pernambuco,
Recife, PE, 50670-901, Brazil}

\author{Mauro Copelli}
\email{mauro.copelli@ufpe.br}
\affiliation{Departamento de F{\'\i}sica, Centro de Ci\^encias Exatas e da Natureza, Universidade Federal de Pernambuco,
Recife, PE, 50670-901, Brazil}

\author{Fernando A. N. Santos}
\email{f.a.nobregasantos@uva.nl}

\affiliation{Dutch Institute for Emergent Phenomena (DIEP), Institute for Advanced Studies, University of Amsterdam, Oude Turfmarkt 147, 1012 GC, Amsterdam, The Netherlands, and Korteweg de Vries Institute for Mathematics, University of Amsterdam, Science Park 105-107, 1098 XG Amsterdam, the Netherlands}

\begin{abstract}
We apply the phenomenological renormalization group to resting-state fMRI time series of brain activity in a large population.
By recursively coarse-graining the data, we compute scaling exponents for the series variance, log probability of silence, and largest covariance eigenvalue. 
The exponents clearly exhibit linear interdependencies, which we derive analytically in a mean-field approach.
We find a significant correlation of exponent values with the gray matter volume and cognitive performance. 
Akin to scaling relations near critical points in thermodynamics, our findings suggest scaling interdependencies are intrinsic to brain organization and may also exist in other~complex~systems.
%
%
%
%
%
%
%
\end{abstract}
\maketitle


Scaling relations near a critical point have a rich history in physics, puzzling researchers for decades until the advent of the renormalization group (RG) theory by Wilson, Kadanoff, Fisher, 
and others~\cite{domb1996critical,wilson1975renormalization, kadanoff2009more}. A key insight from RG theory is that scaling exponents --- quantities describing the scaling behavior of physical observables 
near a phase transition --- are interdependent.  Scaling exponents determine the universality class, so systems with vastly different microscopic details can share the same macroscopic critical behavior.

In complex systems, numerous examples exhibit scale invariance across relevant quantities~\cite{Barabasi99Sci,Sethna2001crackling}. However, identifying how scaling exponents relate to one another has remained challenging in this context. 
While scaling relations are well-established near critical points in equilibrium thermodynamics, complex systems often require a different approach. 
It is frequently not feasible to determine whether a complex system is at a critical point; furthermore, in some conditions systems may display scale invariance without being critical~\cite{Touboul2017, Morrell2021latent,beggs2012being,cantwell2020thresholding}. Nevertheless, it~is still possible to investigate how scaling in different quantities may be connected, regardless of whether the system is in a critical regime.

Renormalization group concepts have been successfully extended to complex systems, with RG-inspired methods applied to neural time series data to uncover underlying scaling behavior~\cite{Meshulam2019coarse,Nicoletti2020scaling,Morales2023scaling,ponce2023critical,castro2024prg,munn2024phylogenetically, meshulam2024review}. In network science, foundational works~\cite{song2005self,song2006origins} have introduced the concept of self-similarity and scaling in networks, providing systematic methods for coarse-graining complex systems and computing scaling exponents analogous to those in standard RG theory. Building on these principles, renormalization flows were explored to gain insights into the hierarchical organization of complex networks~\cite{radicchi2009renormalization,garuccio2023multiscale}. 
Geometric renormalization has also emerged as a powerful approach to uncover multiscale properties, as demonstrated in Refs.~\cite{serrano2009extracting,garcia2018multiscale}, in which the multiscale backbone of weighted networks was extracted and geometric techniques were applied to real networks. 
Similar ideas have been recently applied as well to neural networks~\cite{zheng2020geometric,barjuan2024multiscale}. Very recently, Laplacian renormalization has provided new perspectives on coarse-graining heterogeneous networks, offering methods to study network dynamics and structure at different scales \cite{villegas2023laplacian,loures2023laplacian,nurisso2024higher}. Despite these advances, the quest for a deeper understanding of scaling relations, interdependencies, and universality classes — whether they exist or not in complex systems — remain an open avenue for further research, regardless of the methodology employed.

%



\begin{figure*}[t!]
\centering
 \includegraphics[width=\linewidth]{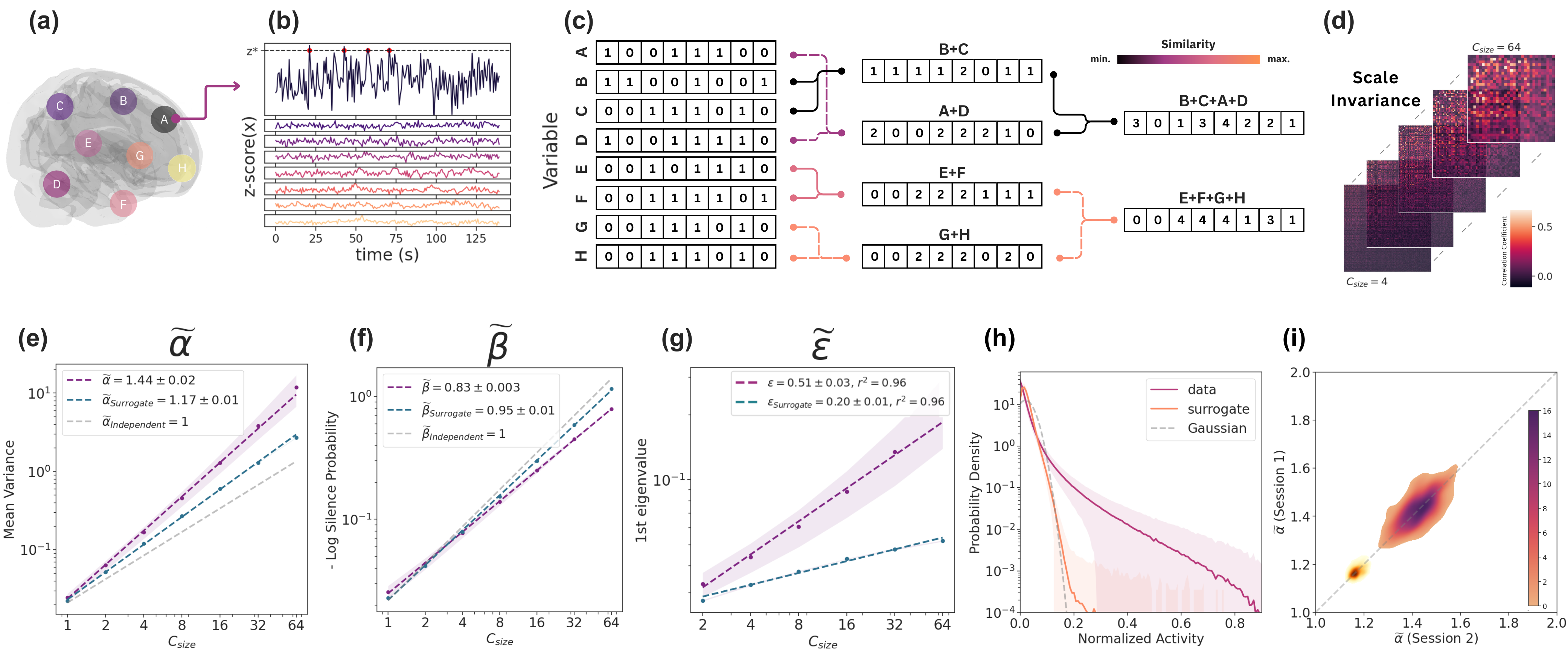}
    \caption{
    (a)-(d) Illustrative scheme of the coarse-graining method. (a)-(b) We first obtain $N = 1014$ variables from BOLD signal time series of ``1000 parcels'' Schaefer Atlas~\cite{schaefer2018local}. Next, for each point in the series we employ a z-score upper threshold~to determine the values of every ROI's binary time series. (c) Then we sum maximally correlated pairs of variables until no single variable is left. (d) We repeat this procedure recursively.  After $k$ successive iterations, the coarse-grained variables correspond to $N/2^k$ clusters of $C_{size} = 2^k = K$ raw signals. 
    In the presence of scale invariance, various statistical features of coarse-grained variables display power-law scaling behavior with the cluster size~$K$, setting the scaling exponents: (e) variance (exponent~$\widetilde{\alpha}$); (f) log probability of complete silence in a cluster ($\widetilde{\beta}$); and (g) largest covariance eigenvalue ($\widetilde{\epsilon}$). Exponents values are also indicated in (e)-(g) for surrogate data and independent variables. (h) The normalized activity distribution approaches a non-Gaussian fixed form under the coarse-graining procedure, as shown for the last iterative step, $C_{size} = 64$. 
    Solid lines represent averages across all HCPYA subjects, while shades depict standard deviations. 
    (i) Density plot of $\widetilde{\alpha}$ in two different sessions for each subject (the~darker the more probable), with the smaller cloud around $\widetilde{\alpha} = 1.2$ corresponding to surrogate~data.
    \label{fig1}
    }
\end{figure*}

In this Letter, we report an instance of such interdependencies. We build upon the phenomenological renormalization group (PRG) approach introduced by Meshulam et al.~\cite{Meshulam2019coarse}, extending it to analyze rs-fMRI data from a large cohort of 714 healthy subjects in the Human Connectome Project~\cite{van2012human}.  
Activity events of brain regions are measured in the form of continuous signals, which can be discretized
into binary (``silent"-``active", Ising-like) variables, $\sigma_i \in \{0,1\}$,  $i = 1,\ldots, N$. 
The idea is to recursively coarse-grain data and observe whether some quantities scale nontrivially  thereupon. 
At each step, the two most correlated variables are merged, pair by pair, until none is left unpaired, see Fig.~\ref{fig1}. 
If $\sigma_i^{(1)}(t) \equiv \sigma_i (t)$ denotes the time series of raw variables,
then after the first PRG iteration one is left with $N/2$ coarse-grained variables $\sigma^{(2)}_i(t)$, defined as 
\begin{equation}
  \sigma^{(2)}_i(t) = \sigma^{(1)}_i(t) + \sigma^{(1)}_{j_*(i)}(t)\; ,
\end{equation}
where $\sigma^{(1)}_{j_*(i)}(t)$ indicates the variable with the highest correlation with respect to $\sigma^{(1)}_{i}(t)$. 
At the $k$-th iteration, one has $N/2^k$ units or ``clusters", each comprising $K=2^k$ raw variables, as illustrated in Fig.~\ref{fig1}. 
From this methodology, we may identify nontrivial scale invariance in the coarse-graining procedure through the change of various statistical quantities as the clusters grow~\cite{Meshulam2019coarse, Nicoletti2020scaling, ponce2023critical, Morales2023scaling, castro2024prg, munn2024phylogenetically}. 
Here we focus on the mean variance $M_2$ of coarse-grained variables, 
log probability of silence~$P_0$ (or effective free energy $F = -\log P_0$), 
and the largest eigenvalue $\lambda_1$ of the covariance matrix (see below), which scale with $K$ respectively with exponents $\widetilde{\alpha}$, $\widetilde{\beta}$ and $\widetilde{\epsilon}$~\cite{Meshulam2019coarse}, 
\begin{eqnarray}
    M_2(K) &\sim & K^{\widetilde{\alpha}}\; , \label{eq2} \\
    F(K) = -\log P_0 (K) &\sim & K^{\widetilde{\beta}}\; , \label{eq3} \\
    \lambda_1 (K) &\sim & K^{\widetilde{\epsilon}} \label{eq4} \; .
\end{eqnarray}
%
\textcolor{black}{In the PRG framework, nontrivial scaling is reflected not only in the values of exponents but also in the non-Gaussian form that the coarse-grained activity distribution approaches. }
In the model-free approach of PRG, such non-Gaussian distributions were found for coarse-grained neuronal activity in different experimental setups~\cite{Meshulam2019coarse, Morales2023scaling, castro2024prg, munn2024phylogenetically}. 
In contrast, in the case of independent coarse-grained variables the joint distribution tends to a Gaussian and the scaling behavior becomes trivial, with exponents $\widetilde{\alpha} = 1$ and $\widetilde{\beta} = 1$. 

\begin{figure*}[t!]
\centering
\includegraphics[width=0.8\linewidth]{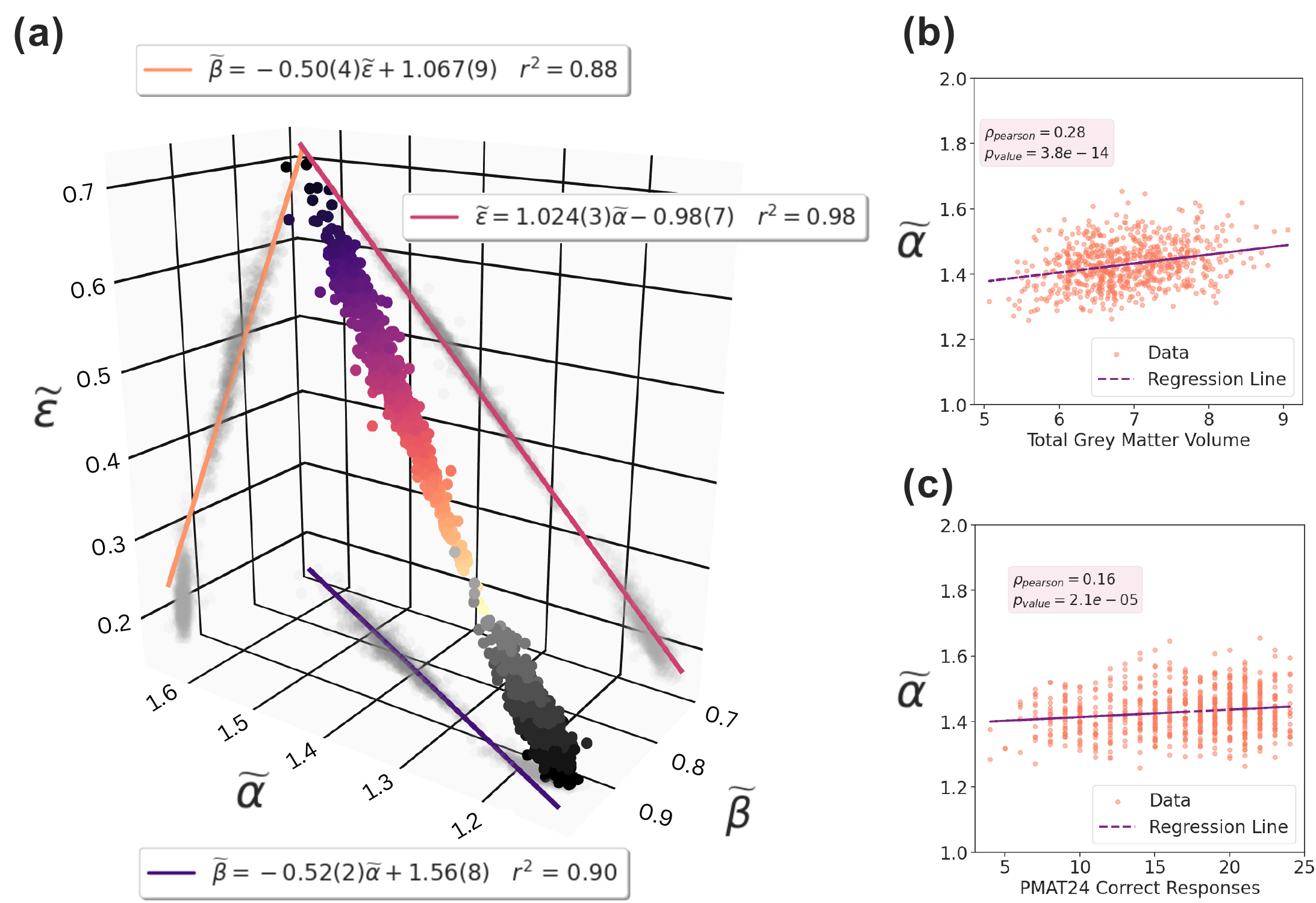}
\caption{
(a) Linear relations between scaling exponents ($\widetilde{\alpha}, \widetilde{\beta}, \widetilde{\epsilon}$). Color circles depict empirical data, color gradient indicates height along the $\widetilde{\epsilon}$-axis, 
and legends display the best linear fit for each projection onto the two-exponent planes, with projected empirical data shown in shaded circles. Points for surrogate data are shown in grey scale. 
(b) Correlation between the exponent~$\widetilde{\alpha}$ and the total grey matter volume (in cm$^3$). (c) Correlation between $\widetilde{\alpha}$ and the number of correct responses in the Penn Progressive Matrices cognitive test (PMAT24).}
\label{fig2}
\end{figure*}

Following the Schaefer atlas~\cite{schaefer2018local}, we parcel each subject's data into 1014 regions of interest (ROIs) shown in Fig.~\ref{fig1}(a). 
Since the rs-fMRI activity distribution is predominantly Gaussian \cite{hindriks2024higher}, we binarize each point in the blood-oxygen level dependent (BOLD) time series of each ROI by using an upper threshold of two standard deviations above the mean~\cite{Tagliazucchi12, ponce2023critical}~[Figs.~\ref{fig1}(b) and~\ref{fig1}(c)].
Surrogate time series can be also obtained by shuffling the Fourier phases~\cite{schreiber2000surrogate}.
We find that the recursive coarse-graining of the PRG procedure indeed yields a non-Gaussian distribution of activity, whereas its surrogate counterpart follows the central limit theorem towards a Gaussian, see Fig.~\ref{fig1}(h).
Consistently, we obtain nontrivial values for the scaling exponents $\widetilde{\alpha}$, $\widetilde{\beta}$ and $\widetilde{\epsilon}$ [Figs.~\ref{fig1}(e)-(g)], 
while exponents for the surrogate data approach their trivial values.
These exponents are relatively stable regarding different sessions of each subject [Fig.~\ref{fig1}(i)].

The relatively large spread in the values of the exponents have not received significant attention in the literature, with prior studies reporting primarily 
only their averages and standard deviations~\cite{Morales2023scaling, ponce2023critical, castro2024prg}. 
However, beyond this analysis and by leveraging a much larger empirical dataset, here we have uncovered a key feature in those variations: they are \textit{strongly} interdependent.
In fact, the exponents $\widetilde{\alpha}$, $\widetilde{\beta}$ and $\widetilde{\epsilon}$ are linearly related, falling remarkably onto a single line in the three-dimensional plot of Fig.~\ref{fig2}(a).
%
Specifically, we observe the following relations: 
\begin{equation}
\begin{aligned}
\widetilde{\beta} &= -0.52 \, \widetilde{\alpha} + 1.56, \quad & R^2 &= 0.90, \\ 
\widetilde{\epsilon} &= 1.02 \, \widetilde{\alpha} + 0.98, \quad & R^2 &= 0.98, \\
\widetilde{\beta} &= -0.50 \, \widetilde{\epsilon} + 1.06, \quad & R^2 &= 0.88, \label{eq5}
\end{aligned}
\end{equation}
which are shown in Fig.~\ref{fig2}(a) as projections onto the two-exponent planes.
These high coefficients of determination ($R^2$) indicate strong linear correlations between the exponents. 
%

To understand the observed relations among the scaling exponents $\widetilde{\alpha}$, $\widetilde{\beta}$, and $\widetilde{\epsilon}$, we derive them analytically in a mean-field approach. 
%
We start by considering the joint probability distribution of the binary variables $\{\sigma_i\}$ within a cluster of size $K$, modeled using pairwise interactions akin to an Ising model,  
\begin{equation} P_K(\{\sigma_i\}) = \frac{1}{Z_K} \exp\left( \sum_{i=1}^K h_i \sigma_i + \frac{1}{2} \sum_{i=1}^K \sum_{j=1}^K J_{ij} \sigma_i \sigma_j \right),
\label{eq6}
\end{equation} 
where $Z_K$ is the normalization constant, $h_i$ are local fields, and $J_{ij}$ are interaction strengths between variables.
We note that Eq.~(\ref{eq6}) has the form of the distribution resulting from pairwise maximum entropy models, which have been successfully applied to describe the collective behavior of neural networks~\cite{new1,schneidman2006weak}.
In this context, $\{J_{ij}\}$ and $\{h_i\}$ are determined so that the
means~$\langle \sigma_i \rangle$ and covariances $C_{ij} = \langle \sigma_i \sigma_j \rangle - \langle \sigma_i \rangle \langle \sigma_j \rangle$ calculated from $P_K$ agree with the experimental data. 
By writing $P_K = \exp (-E_K(\{\sigma_i\}))/Z_K$, with the ``energy" $E_K$ of state $\{\sigma_i\}$, an effective free energy $F(K) = - \log Z_K = -\log P_0$ can be identified from~(\ref{eq6}), where the probability of silence $P_0$ corresponds to the state~${\{ \sigma_i = 0 \}}$.  
It has been shown~\cite{Meshulam2019coarse,new3} that the discretization procedure on the variables $\{ \sigma_i \}$ preserves the structure of correlations and statistical properties of the system, with Eq.~(\ref{eq6}) holding as well for continuous variables~$\{ \sigma_i \}$.

By identifying $h_i = -\sum_{j=1}^K (C^{-1})_{ij} \langle \sigma_j \rangle$ and $J_{ij} = (C^{-1})_{ij}$ for continuous variables, Eq.~(\ref{eq6}) can be cast into the form of the multivariate normal distribution, 
\begin{equation} P_K(\mathbf{x}) = \frac{1}{(2\pi)^{K/2} |\det(C)|^{1/2}} \exp\left( -\frac{1}{2} \mathbf{x}^\mathrm{T} C^{-1} \mathbf{x} \right), 
\label{eq7}
\end{equation} 
where $\mathbf{x} = \boldsymbol{\sigma} - \langle \boldsymbol{\sigma} \rangle$, with $\boldsymbol{\sigma} = (\sigma_1, \dots, \sigma_K)$, and $C$ as the covariance matrix. 
As stated by the multivariate central limit theorem~\cite{new2}, the joint probability density of scaled sums of multiply correlated $K$-dimensional variables $\boldsymbol{\sigma}$ with finite variances $\{C_{ii}\}$ converges to the multivariate normal distribution~(\ref{eq7}).  
Indeed, we can infer from the fMRI data that Eq.~(\ref{eq7}) (or equivalently Eq.~(\ref{eq6})) is a suitable model for $P_K$ as the empirical variables are widely correlated and display finite variances~\cite{hindriks2024higher}. 
%
%

%

To establish a connection between the observables with scaling behavior given in Eqs.~(\ref{eq2})-(\ref{eq4}) and obtain scaling relations between the exponents $\widetilde{\alpha}$, $\widetilde{\beta}$, and~$\widetilde{\epsilon}$, 
%
we consider a mean-field approach analogous to the Curie-Weiss model~\cite{domb1996critical}, in which $C_{ii} = v$ and $C_{ij} = c$, $i \neq j$, in Eq.~(\ref{eq7}).
This approximation assumes equal interactions among all variables, simplifying the covariance structure while capturing the essential features necessary for the analysis. 
In fact, though some variability does exist in the covariance profile of rs-fMR data, inference from the empirical data series indicates that a fraction of 10\% of the interval of $C_{ij}$-values concentrate about 60\% of all covariances, with the probability decreasing exponentially for covariance values progressively further away from this range. 
As for the variances $C_{ii}$ with nearly Gaussian distribution, we notice that about 68\% of empirical $C_{ii}$-values lie in an interval of two standard deviations around the mean. 
Note that the central limit theorem drives the distributions of correlations in the rs-fMR data.

The covariance matrix in this approach reads
\begin{equation} C = (v - c) I_K + c \mathbf{1}_K \mathbf{1}_K^\mathrm{T}, \end{equation} 
where $I_K$ is the $K \times K$ identity matrix and $\mathbf{1}_K$ is a $K$-dimensional vector of ones.
We note that $\det(C) = {(v-c)^{K-1}[v + (K - 1)c]}$. 
The inverse of $C$ is given~by 
\begin{equation} C^{-1} = \frac{1}{(v - c)} \left( I_K - \frac{c}{[v + (K - 1)c]} \mathbf{1}_K \mathbf{1}_K^\mathrm{T} \right). \end{equation}
The mean energy, 
$E (K) \equiv \langle E_K(\{\sigma_i\}) \rangle$, 
can be computed from the results above, yielding
\begin{equation} 
E (K) = \frac{v^{1/2} K}{2 [v + (K - 1)c]^{1/2}} \prod_{n=0}^{K-2} \left( \frac{v + c + n c}{v + n c} \right)^{1/2}. 
\label{eq12}
\end{equation}
Moreover, as the PRG procedure recursively assigns each coarse-grained variable to a sum over $K$ original variables, we calculate the variance of~the~sum~${S_\sigma (K) = \sum_{i=1}^K \sigma_i}$, leading to 
%
\begin{equation} 
M_2 (K) = \operatorname{Var}(S_\sigma) = v K + c K (K - 1). 
\label{eq11}
\end{equation}
%
%
For large $K$, the fractions inside the product in Eq.~(\ref{eq12}) tend to unity, 
and so we combine with Eq.~(\ref{eq11}) to find 
\begin{equation} 
E \sim \frac{K^{3/2}}{M_2^{1/2}} \sim K^{(3 - \widetilde{\alpha})/2},  \end{equation}
where we have used $M_2 \sim K^{\widetilde{\alpha}}$, Eq.~(\ref{eq2}). 
%
%

We remark that under the PRG approach applied to neural networks~\cite{new1,new3} the entropy $S \to E$ in the thermodynamic limit $N \to \infty$ of the number of neurons. 
However, for finite~$N$ numerical and empirical PRG studies~\cite{new1} have pointed that the contribution of the entropy to the free energy $F$ is appreciably smaller than that of the mean energy in the large-$E$ regime. 
So in this regime we also obtain $F(K) \sim K^{(3 - \widetilde{\alpha})/2}$, which from the scaling form $F(K) \sim K^{\widetilde{\beta}}$, Eq.~(\ref{eq3}), allows to identify the scaling relation between the exponents~$\widetilde{\alpha}$~and~$\widetilde{\beta}$,
\begin{equation} \widetilde{\beta} = \frac{3 - \widetilde{\alpha}}{2}.
\label{eq13}
\end{equation}
%


We next turn to the exponent $\widetilde{\epsilon}$ of the largest eigenvalue $\lambda_1$ of the covariance matrix~$C$, which reads
%
\begin{equation} 
\lambda_1(K) = v + c(K - 1), \end{equation} 
so that $\lambda_1(K) = M_2 / K$. 
From $\lambda_1 (K) \sim  K^{\widetilde{\epsilon}}$, Eq.~(\ref{eq4}), the following relation is obtained,
\begin{equation} \widetilde{\epsilon} = \widetilde{\alpha} - 1.
\label{eq15}
\end{equation}


These analytical results agree quite well with the empirical data, as observed in Fig.~\ref{fig2}(a) and also by comparing Eqs.~(\ref{eq13}) and (\ref{eq15}) with the empirical fit expressions in~Eq.~(\ref{eq5}). 
%

Notably, the scaling relations (\ref{eq13}) and (\ref{eq15}) also accommodate well both the limit case of independent variables $(\widetilde{\alpha} = 1, \widetilde{\beta} = 1, \widetilde{\epsilon} = 0)$ and the exponents from surrogate data $(\widetilde{\alpha} \approx 1.17, \widetilde{\beta} \approx 0.95, \widetilde{\epsilon} = 0.20)$, see  Figs.~\ref{fig1}(e)-(g) \textcolor{black}{and the lowest region of data points in Fig.~\ref{fig2}.}

Importantly, we have also found that variations in the exponents are intrinsically connected with differences in anatomical and behavioral traits across subjects, as~shown in~Figs.~\ref{fig2}(b) and \ref{fig2}(c). 
For instance, the exponent $\widetilde{\alpha}$ (and, therefore, any of the three exponents due to their interrelations) is significantly correlated with the total gray matter volume (see Fig.\ref{fig2}(b), with Pearson coefficient $\rho = 0.28$ and $p$-value $p = 3.8 \times 10^{-14}$)~and to~the number of correct responses in the Penn Progressive Matrices cognitive test (PMAT24) (Fig.~\ref{fig2}(c), $\rho = 0.16$, $p = 2.1 \times 10^{-5}$). This association aligns with the striking notion that variability in the brain's scaling exponents reflects variability in 
behavioral and anatomical scores~\cite{cocchi2017criticality,zimmern2020brain}.


In conclusion, the uncovering of interdependent scaling relations between exponents driving the scaling behavior of relevant quantities in rs-fMRI datasets point to an intriguing level of brain's organization. 
While it might be premature to claim that the brain's scaling behavior belongs to some universality class, the emergence of such interdependencies hints at universal organizing principles in neural systems. 
One limitation of our approach is considering all variables with homogeneous variance and covariance, which simplifies the mathematics but possibly overlooks the inherent variability in brain activity. 
Also, while the central limit theorem concurs with a mean-field approach to rs-fMRI data, deviations from Gaussianity may be relevant especially at faster time scales, such as in spiking, EEG or MEG data.
So extending this framework to other models, data types and  renormalization approaches could reveal further hidden features of neural networks' collective behavior. 
%

%
Finally, given that our approach relies solely on time series data analysis, it also opens the possibility of finding scaling interdependencies in other complex systems exhibiting multiscale dynamics. 
%
%
Exploring these systems could advance the general understanding of universality and pave the way towards identifying key underlying mechanisms governing complex system behavior.
We thank Floris Benjamin Tijhuis for preprocessing of the HCP data \cite{tijhuis2021hcp_rfMRI}.  Data were provided by the Human Connectome Project (HCP), funded by NIDCR, NIMH, and NINDS, and shared by the Laboratory of Neuro Imaging, University of Southern California. We also thank the Brazilian agencies CAPES, CNPq (grants 308840/2023-2 (EPR) and 308703/2022-7 (MC)), and FACEPE.  
DMC gratefully acknowledges a visiting scholarship of the Dutch Institute for Emergent Phenomena (University of Amsterdam), during which part of this work was developed. 


%

\end{document}